\documentclass[conference]{IEEEtran}
\usepackage{graphicx}

\begin{document}
\setlength{\columnsep}{0.2in}
\title{
Exploiting Satellite Broadcast despite HTTPS
}

\author{\IEEEauthorblockN{Nikos Fotiou, Vasilios A Siris,\\ George C. Polyzos 
}
\IEEEauthorblockA{Mobile Multimedia Laboratory,\\
Department of Informatics,\\
Athens University of Economics and Business\\
Email:\{fotiou, vsiris, polyzos\}@aueb.gr}
\and
\IEEEauthorblockN{Mario Marchese, Franco Davoli 
}
\IEEEauthorblockA{Satellite Communications and Networking \\Laboratory\\
DITEN Department\\
University of Genoa\\
Email:\{firsname.lastname\}@unige.it}
\and
\IEEEauthorblockN{Luca Boero, 
}
\IEEEauthorblockA{
Rulex Inc.,\\
Email:l.boero@rulex.ai}
}

\maketitle

\begin{abstract}
HTTPS enhances end-user privacy and is often preferred or enforced by over-the-top content providers,
but renders inoperable all intermediate network functions operating above the transport layer,
including caching, content/protocol optimization, and security filtering tools. 
These functions are crucial for the optimization of integrated satellite-terrestrial networks.
Additionally, due to the use of end-to-end and per-session encryption keys, the advantages 
of a satellite's wide-area broadcasting capabilities are limited or even negated completely.
This paper investigates two solutions for authorized TLS interception that involve TLS splitting.
We present how these solutions can be incorporated into integrated satellite-terrestrial networks
and we discuss their trade-offs in terms of deployment, performance, and 
    privacy.
Furthermore, we design a solution that leverages satellite broadcast transmission even in 
the presence of TLS (i.e. with the use of HTTPS) by exploiting application layer encryption
in the path between the satellite terminal and the TLS server.
Our findings indicate that even if no other operation than TLS splitting is performed,
TLS handshake time, which involves roundtrips through possibly a Geosynchronous satellite,
can be reduced by up to 94\%. Moreover, by combining an application layer encryption solution with TLS splitting, 
broadcast transmissions can be exploited 
    as well as proactive caching, content pushing,
request aggregation, and other optimizations.

\end{abstract}

\IEEEpeerreviewmaketitle

\section{Introduction}
It is evident that the Web is becoming encrypted. Initiatives such as 
    {\em Let's Encrypt}
allow hassle free and 
    no-cost HTTPS services.
Therefore, it comes to no surprise that according to Google, Chrome users 
spend two-thirds of their time in HTTPS pages~\cite{goo}.  
Similarly, HTTPS pages receive higher ranking by search engines and
browsers already mark plain HTTP pages as non-secure. At the same time, 
privacy concerns have led to the use of TLS even for transferring DRM protected 
content (e.g., the case of Netflix \cite{net}). 
    Although HTTPS improves significantly end-user security and privacy, 
it comes with a cost: it prevents in-network functions, such as network-based
security services, application-level gateways and 
    fine-level differentiation of services, session controllers,
transcoders, proxies, and caches. 

In-network content manipulation is not uncommon in wireless and mobile 
networks and it is mainly used for optimizing network performance (as perceived
by both the network operators and the end-users). Naylor et al.~\cite{Nay2014} 
reported that a transparent proxy used by a major European 
mobile carrier, serving more than 20 million subscribers, contributes to 
a 2TB/day decrease of upstream traffic using caching and 28.5\% decrease 
of last-mile downstream traffic using compression. Woo et al.~\cite{Woo2013} reported 
that standard Web caching can reduce download 
bandwidth consumption up to 27.1\%. Sivakumar et al.~\cite{Siv2014} 
developed a proxy--code named PARCEL--for mobile networks that pre-fetches 
and pre-processes Web content: browsing 34 Web pages from the top 500
Alexa global pages using PARCEL resulted in
    a
49.6\% reduction in page 
load time and 65\% reduction in energy consumption. 
Similarly, various publications (e.g.,~\cite{Sek2011}, ~\cite{Sher2012}) 
report that the number of middleboxes that manipulate network traffic in big enterprise networks is almost 
equal to the number of L2 switches and L3 routers. All these functions
can be used for improving the performance of integrated satellite-terrestrial networks,
as well as for decreasing the latency introduced by the satellite part of
those networks. However, the use of TLS/HTTPS affects all of them. Additionally,
end-to-end encryption, as well as the use of per-session encryption keys 
(all imposed by TLS) render broadcast communication useless, since a 
content encrypted for a specific session is just ``junk'' for all other
sessions. This has a huge impact to satellite-terrestrial communications,
where broadcast is widely used for delivering content. 

End-to-end encryption has been a problem for Content Distribution Networks (CDNs) for a long time now.
For this reason CDN providers are using solutions, such as \emph{Custom Certificates} 
and \emph{Certificate sharing}~\cite{Lia2014}
    that allow them to intercept
TLS connections.
However, these solutions require a long-term trust relationship between the
content owner and the CDN provider. Furthermore, these solutions assume that
CDN providers are trusted to stop intercepting TLS traffic whenever requested
by content owners. These requirements can be hardly satisfied when it comes
to TLS interception in an integrated satellite-terrestrial network 
since satellite terminals (i.e., the location where TLS split should take place)
are not so well protected (compared to a CDN node),
and their operators cannot always be trusted by content owners. Therefore,
using these solutions in such an architecture would create intolerable security
and privacy risks.

In this paper we leverage two solutions that allow authorized TLS splitting by in-network
devices and we use them in the context of integrated satellite-terrestrial networks. These
solutions, namely Keyless TLS~\cite{Ste2015} and DANE with delegation semantics~\cite{Lia2014}, 
allow a content owner to temporarily authorize a device to (lawfully) 
intercept a specific TLS session. In essence, these solutions \emph{split}
a TLS connection into two parts: one between the TLS client and the device
and another between the device and the TLS server; we use these solutions
in order to split a TLS connection at the satellite terminals. Additionally,
we leverage the fact that due to this split in the TLS connection, the algorithms and protocols  
for securing the path between the satellite terminal and the TLS server
are hidden from the TLS client; hence, we can
apply application layer security solutions in that part: 
these solutions, if configured properly, do not impede satellite's
broadcasting capabilities. The contributions of our work presented in this paper are:
\begin{itemize}
\item We design an integrated satellite-terrestrial network architecture which
    incorporates TLS splitting mechanisms.
\item We analyze the performance and security properties of our architecture
    through analysis and simulation.
\item We design an extension to our architecture that uses application layer
encryption in the path between the satellite terminal and the service provider,
enabling solutions that take advantage of the broadcast capabilities of the satellite
network.
\end{itemize}

The structure of the remainder of this paper is as follows. In Section 2 we present background
information and we introduce the selected
solutions. In Section 3 we present how the selected solutions can be used with
integrated satellite-terrestrial networks and we evaluate our approach in Section 4.
Finally, in Section 5 we provide our conclusions and plans for future work.

\section{Background}

    \subsection{Transport Layer Security}
    The primary goal of the Transport Layer Security (TLS) protocol is to provide privacy and data integrity 
between two communicating endpoints; a client and a server~\cite{Die2008}. TLS enables the establishment of 
a secure connection that protects the confidentiality and the integrity
of the transmitted data.
TLS is composed of two protocols: the Handshake Protocol and the Record Protocol. 
    The  Handshake Protocol allows a client and a server to authenticate each 
other and to negotiate security algorithms, as well as the 
corresponding cryptographic keys. The TLS Record Protocol is then used for 
securing application layer data using the agreed keys and algorithms. 
The Handshake protocol is critical when it comes to TLS splitting; for this reason we provide 
some more details about it next. 

A TLS Handshake is completed in 3 steps. The first 
step is the cipher suite negotiation. In this step the client and server exchange ``Hello'' messages 
and choose the cipher suite that will be used throughout a session. The second step 
is authentication. In TLS, a server proves its identity to the client and a client may 
also prove its identity to the server. Digital certificates (and their corresponding private keys) are the basis of 
this authentication whereas the exact method used for authentication is determined by the cipher 
suite negotiated. In any case, the authentication process involves the private key of a public-private 
key pair, owned by the authenticating entity. The final step is key exchange where the client and server exchange random 
numbers which combined 
with additional data permit the secure calculation of the
session-specific shared  keys. 

\subsection{Keyless TLS}
\begin{figure}
\centering
\includegraphics[width=0.90\linewidth]{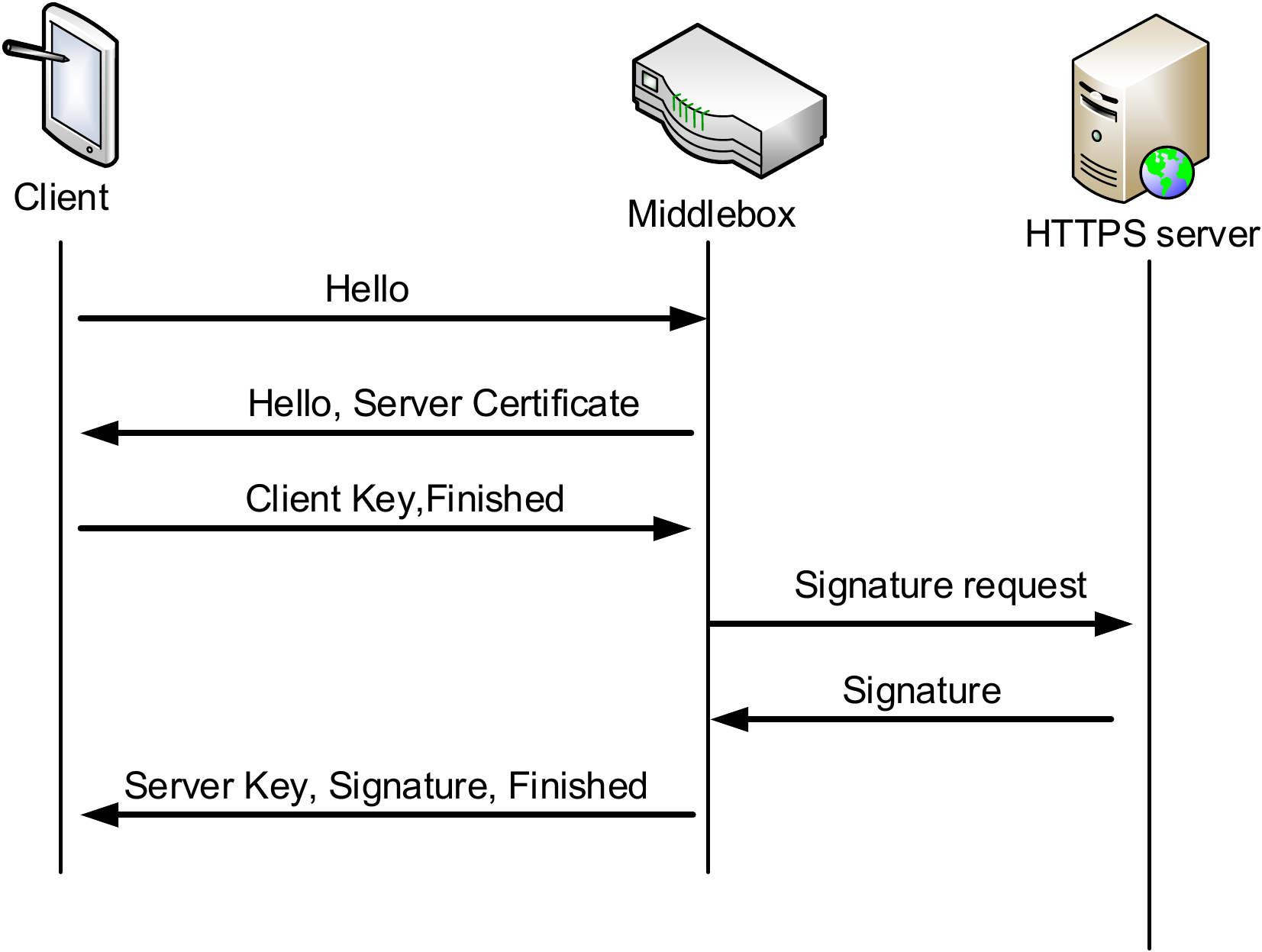}
\caption{Keyless TLS.}
\label{keyless}
\end{figure}

Keyless TLS allows authorized devices to intervene in a secured connection, but without having 
access to any private key, hence they cannot be authenticated without the ``help'' of the TLS server.  
In particular, the intercepting device performs 
the TLS handshake and responds on behalf of the server, but all handshake operations requiring 
the private key (of the server) are relayed to the server over a dedicated secured channel: the server authenticates the 
intercepting device, 
performs the private key operations, and returns the result through the same secured channel.
This process is illustrated in Figure~\ref{keyless}.
Cloudflare~\cite{Sou2014} and Akamai~\cite{ger2017} are offering keyless TLS as a service.

Keyless TLS has two significant advantages: intercepting devices do not learn private keys  
and TLS servers participate in all session establishments, hence they can prevent at any time 
a device from intervening in an encrypted connection. Moreover, keyless TLS requires no modification
 to TLS clients. Of course, keyless TLS does not come without disadvantages. 
Its main drawback is the weakening of end-to-end security since, in reality, 
a TLS connection is split into two independent connections that involve two different TLS handshakes; 
one of them may result in a weak cipher which reduces the security of the end-to-end connection. 
According to a study~\cite{Lia2014} performed across many major CDN services that use similar techniques this phenomenon is common 
and there were even cases where the connection between the intercepting CDN node and the server was not secured 
at all: sensitive information was transferred using plain HTTP, yet clients were under the impression 
they were using an end-to-end encrypted connection.
\subsection{DANE with delegation semantics}
\begin{figure}
\centering
\includegraphics[width=0.90\linewidth]{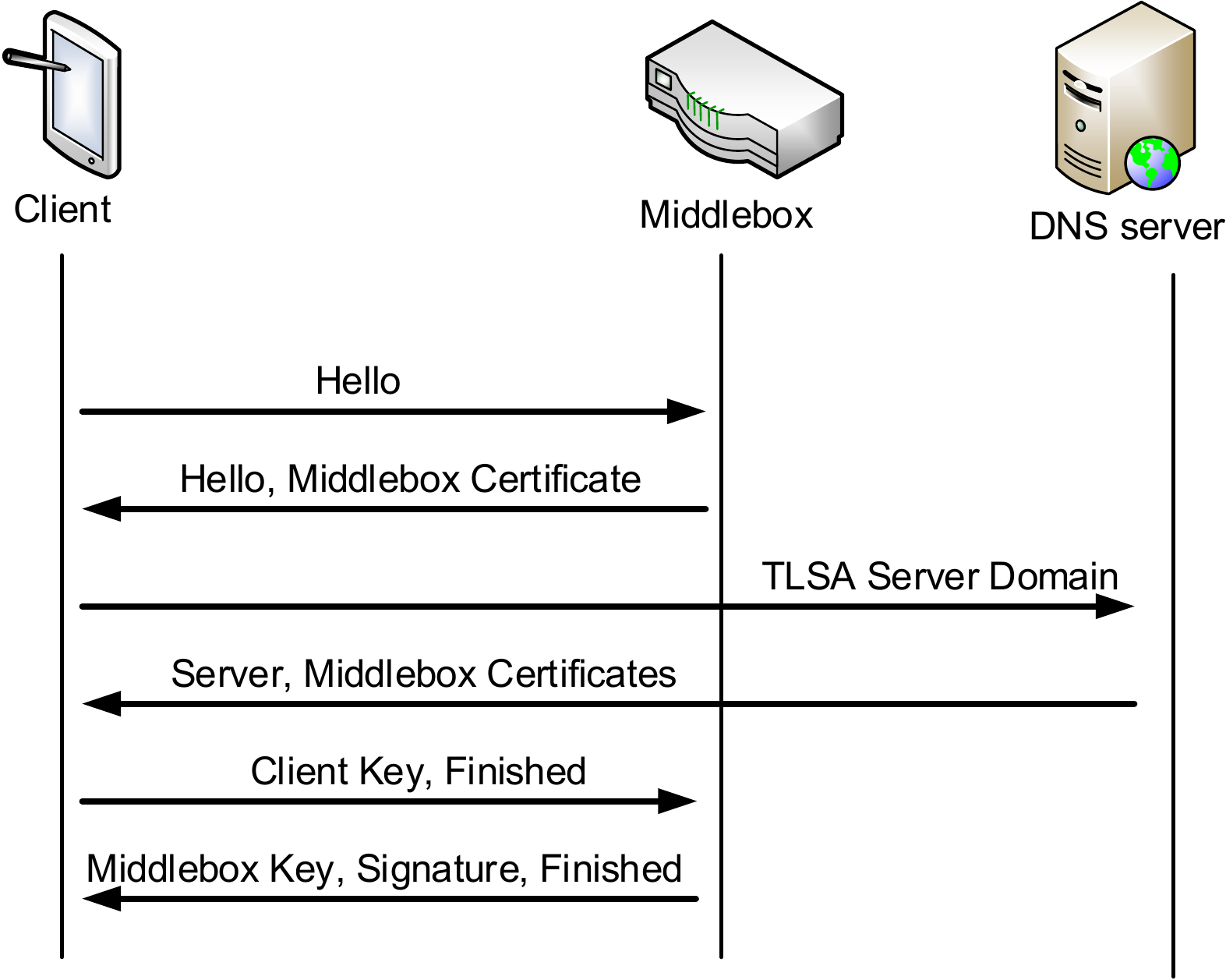}
\caption{DANE with delegation semantics.}
\label{dane}
\end{figure}

DNS-Based Authentication of Named Entities (DANE) (originally proposed in RFC 6698~\cite{Sch2012} and further
refined in RFC 7671~\cite{Har2015}), allows binding a domain name to a certificate. This binding 
is implemented by including certificates in DNS records.\footnote{
    In particular digitally signed hashes of the certificates.} 
In particular, a special
type of DNS record, referred to as TLSA DNS, is used for associating certificates 
with domain names, allowing DANE-enabled TLS clients to validate TLS server certificates. 
    The resolution of TLSA records is secured using DNSSEC. 
DANE with delegation semantics~\cite{Lia2014} leverages DANE, allowing servers to add intermediate certificates
to their TLSA records, which can be used by intercepting devices. This way a client may obtain 
a list of certificates that can be used for 
a particular TLS connection. Figure~\ref{dane} illustrates an example of a TLS handshake interception
supported by DANE. As it can be observed, the intercepting device responds to the 
    client ``Hello'' message with its own certificate (as opposed to the Keyless TLS case, where
    the intercepting device responds with the certificate of the server). Subsequently, the client performs
a  TLSA record resolution and validates that the received certificate is ``pinned''
to the server's domain name, therefore it is approved. Finally, the client proceeds with the subsequent TLS handshake
messages. 

    With this solution, certificate revocation is completely 
controlled by the origin server and can be performed by simply altering the corresponding TLSA record. 
Furthermore, an interesting 
property of this solution is that a handshake can be completed without
any involvement of the original TLS server.
Deploying this solution requires DNSSEC along with modifications to the certificate 
validation process performed by TLS clients.  
    Of course, this approach suffers from the DNSSEC inherent problems, 
for example, an attacker may replay a TLSA record response related to a certificate that is 
not valid any more. 
Overhead is also added due to the extra DNS round trip. 
\begin{figure*}
\centering
\includegraphics[width=0.68\linewidth]{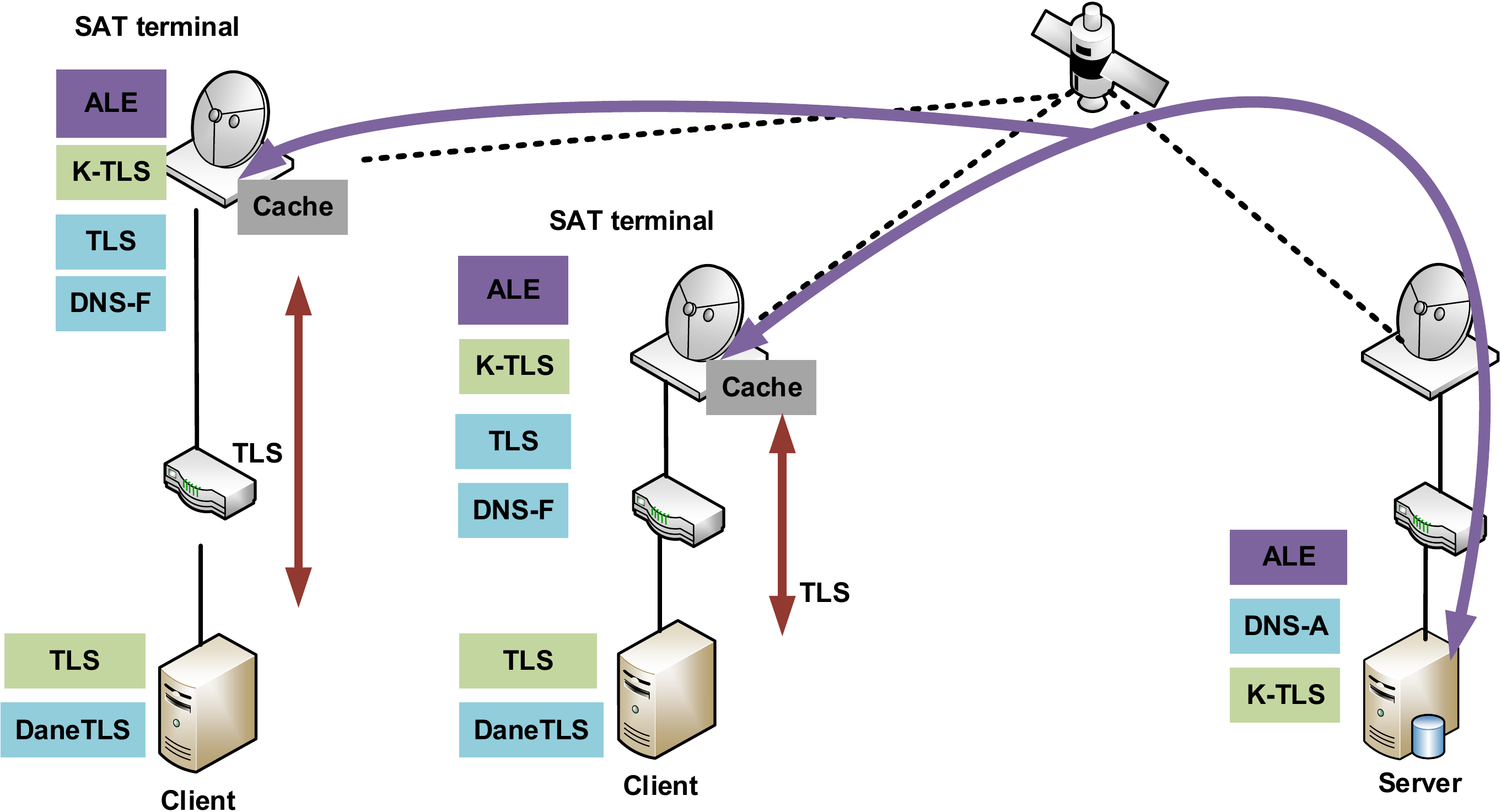}
\caption{TLS interception combined with application layer encryption.}
\label{arch}
\end{figure*}

\section{TLS interception for Integrated Satellite-Terrestrial Networks}
In this section we describe our solution for supporting TLS interception
in integrated satellite-terrestrial networks. We consider a typical 
integrated satellite-terrestrial network architecture
where a satellite terminal and a satellite gateway are responsible for connecting
client applications with server applications over a satellite network (see for example Figure~\ref{arch}).
    In the following we consider that clients and servers wish to communicate 
over TLS and we propose two approaches that allow terminals to perform TLS interception so as (i)
to accelerate the TLS handshake--which under normal circumstances has 
to be performed over the satellite network, and (ii) to perform content transmission 
optimizations (e.g., content caching, aggregation of content requests, etc.).

    Our two approaches are based on the TLS interception solutions presented in the
previous sections. With both solutions, whenever a terminal receives a client 
``Hello'' TLS message (i.e., the first message of the TLS handshake)
it should be able to tell
(i) if it is allowed to intercept this TLS handshake, and
(ii) which certificate to use.
Both these problems can be solved using the
Server Name Indication TLS extension~\cite{Eas2011}.
With this extension a TLS client includes
in its ``Hello'' message the domain name of the service with
which it wants to interact. We consider that satellite terminals are pre-configured with the
domain names of the services they are authorized to intercept,
therefore they can examine the client ``Hello'' message  and decide
whether to intercept the handshake or to forward the  message to the intended 
recipient.\footnote{This extension is well supported by all browsers since it is used for connecting to a TLS server hosted in a shared (e.g., Cloud) environment.}


\subsection{Integration with Keyless TLS}
Keyless TLS is integrated in our  architecture
by implementing the Keyless TLS protocol at the terminals and the TLS servers;
no modification is required to TLS clients
(these components are depicted with a green box in Figure~\ref{arch}).
Furthermore, a secure communication channel is established between the terminal and the server.
This channel, which in our implementation is secured using TLS, 
can only be used by terminals authorized to intervene in a TLS handshake
and its purpose is to protect the confidentiality of the Keyless TLS specific 
messages transmitted between the terminal and the server.
In order to assure that only authorized terminals can access this channel
TLS client authentication with certificates~\cite{Sou2014}
is used i.e., for each authorized terminal,
a TLS server generates a certificate, which is installed using out-of-band 
mechanisms and it is used by the terminal to authenticate
itself to the server when setting up the channel. The channel setup 
takes place only once and the same channel is used  for forwarding Keyless TLS specific 
messages for all subsequent handshakes intercepted by the same terminal. With all these components in place, a TLS client
initiates the TLS handshake, which is intercepted by the terminal and whenever the terminal requires
to perform an operation using the private key of the server, it sends all
necessary information to the server through the secure channel; then the server
performs the necessary actions and responds back to the terminal, again through the secure channel. 

An interesting property of Keyless TLS is that the server participates in all TLS handshakes.
Although, this property enhances the security of the scheme
    (we discuss security properties in the following sections), it adds latency
since the satellite network has to be used once per TLS handshake.
In some cases, it is possible to compensate for this delay by ``abusing'' the SNI TLS extension:
since the domain of the service in which the client is interested in is known,
it may be possible to \emph{push} content to a terminal together with the first Keyless TLS message.
Then, when requested, this content can be served to a client directly by the terminal (which in this case
    acts as a transparent cache),
    hence the satellite link does not have to be used.     

\subsection{Integration with DANE with delegation semantics}
For the integration of DANE with delegation semantics, clients implement TLS with DANE assisted certificate
verification, i.e., they are able to verify the validity of a digital
certificate by retrieving the corresponding TLSA DNS record (using DNSSEC). The components of this
solution are depicted with a blue box in Figure~\ref{arch}. With this approach
a secure communication channel between the terminal and the server is not
required in order to complete the TLS handshake.

As already discussed, during the TLS handshake clients should perform a
DNS resolution in order to validate the certificate of the terminal. In the
general case, this resolution will cross the satellite network. However,
this can be easily mitigated by installing DNS forwarders in satellite terminals
and by configuring clients to use these forwarders as the default DNS server.
The forwarders, which should implement the DNSSEC validation processes, 
can then cache the corresponding DNS replies. Furthermore, and since
these replies concern the terminal operation, the forwarders have incentives
to periodically perform DNS requests to the authoritative DNS server (even if they
are not instructed by a client) in order to keep their cache fresh and up to date.

\subsection{Combination with application layer encryption}
With both approaches the client ends up establishing a TLS connection with 
the satellite terminal (the difference lies in the fact that with Keyless TLS
the client ``thinks'' that it is communicating to the server, whereas with DANE with delegation semantics
the client knows that it communicates with the terminal, acting on behalf
    of the server). Clients are oblivious to the security mechanisms
used for securing the communication between the terminal and the server.
A typical approach would have been to use another TLS connection between 
these two entities; alternatively an application layer encryption mechanism can be used
    so that solutions that leverage a satellite's wide area broadcasting capabilities will be able to function properly.
The components of this approach are illustrated with purple colors in Figure~\ref{arch}. 

Our application encryption approach assumes that servers and (authorized)
terminals share a secret key. This key is used for periodically exchanging
content encryption keys. The latter keys are used for encrypting transmitted 
content and are common for all terminals. This is a typical mechanism used for 
    broadcasting protected content over satellites, e.g., as used by the 
    Conditional Access system of the  Digital Video Broadcasting (DVB) standard.
However, the content encryption key is not related to the encryption key
used in the TLS connection between the client and the terminal. For this reason
    the terminals have to decrypt the received items and re-encrypt them with the
corresponding TLS key.   

\section{Evaluation}
\subsection{TLS Handshake speed improvements}
\begin{figure}
\centering
\includegraphics[width=0.90\linewidth]{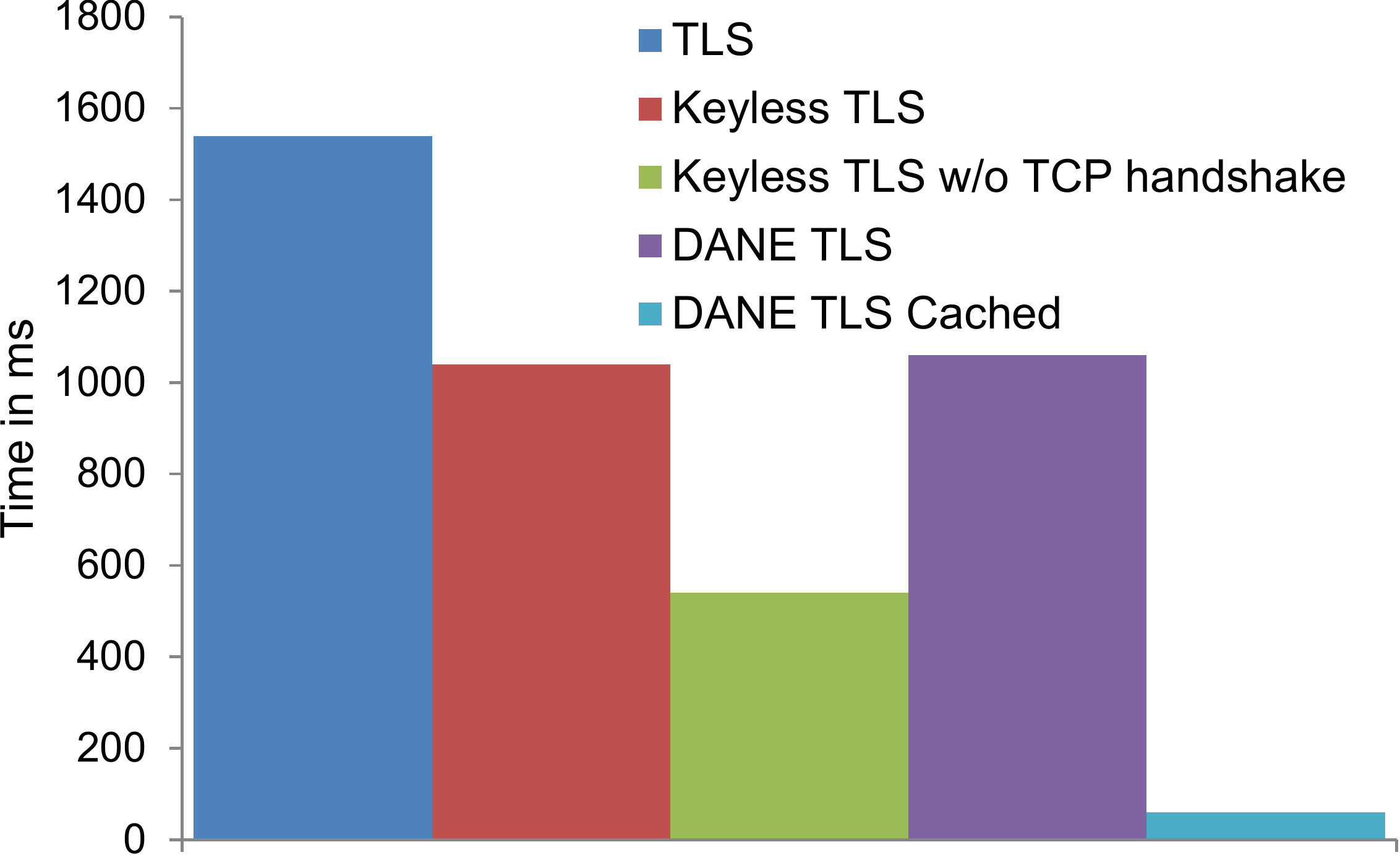}
\caption{Time required to complete a TLS handshake.}
\label{evaluation}
\end{figure}

    Compared to ``vanilla'' TLS, the discussed TLS interception solutions require
fewer message exchanges over the satellite network. Therefore, even if the satellite terminals 
implement no additional content transmission optimization function, significant gains can be achieved. Figure~\ref{evaluation}
illustrates the time required to complete a TLS handshake in a network where
a terminal and a gateway are connected through a Geosynchronous satellite. 
The roundtrip delay of the path between the terminal and the gateway
has been calculated to be 500 ms using the OpenSAND satellite network emulator~\cite{open}. 
Furthermore, the roundtrip delay of a terrestrial path has been set to 
20 ms. It can be observed that when a TLS interception solution is used, 
    the TLS handshake is completed much faster. Especially, when DANE with delegation semantics
is used combined with cached DNS records, the satellite network does not have
to be used during a TLS handshake, hence the TLS handshake time is reduced by 94\%. 

\subsection{Content transmission optimizations}
Here we discuss some content transmission optimizations that become possible
if the application layer encryption approach is used. 

The application layer encryption solution used in our system is 
Encrypted Content-Encoding for HTTP, specified in RFC 8188~\cite{Tho2017}.
This solution enables HTTP messages
(request or response) to be encrypted using a 
symmetric encryption key distributed using out-of-band mechanisms. In the following
we give an example that illustrates how the combination of application
layer encryption with TLS interception can benefit integrated satellite-terrestrial
networks. In this example we assume the architecture of Figure~\ref{arch}. In 
this architecture the terminals and the servers share a symmetric encryption key.
Furthermore, the terminals include a cache and support TLS interception.

\begin{figure}
\centering
\includegraphics[width=0.90\linewidth]{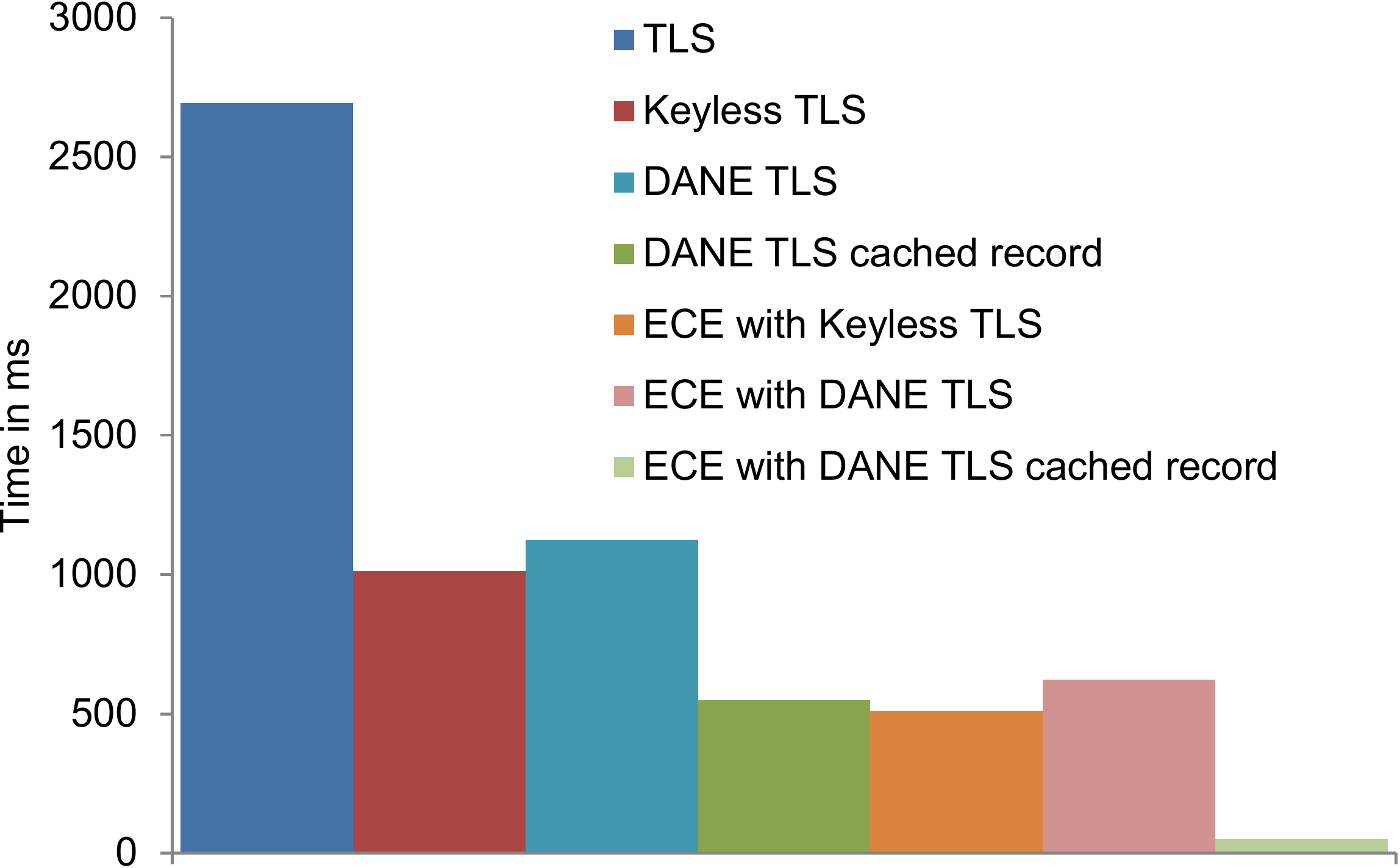}
    \caption{Time to load a Web page using HTTPS.}
\label{results2}
\end{figure}

    Suppose an end-user requests a piece of static content over TLS (e.g., a video file), 
its terminal intercepts the TLS session and forwards the request to the server.
The server encrypts the requested content using the shared encryption key and transmits it over plain HTTP.
The terminal re-encrypts  the received content (to match the TLS encryption key) and transmits it back to the client. 
    All other terminals can cache this piece of content (since it is being broadcast). Then, if
another client, connected to (the same or) another terminal, requests the same piece of content (using TLS),
its terminal can re-encrypt the cached content using the established (between them) TLS 
key and send it back to the client.
Therefore, although there are two clients in this case, connected to (potentially) different 
terminals, requesting the same piece of content, 
    there is a single content request, as well as, a single content transmission
    (the bulk of the data) traversing the satellite link. 
There are of course two separate (potentially asynchronous) data transmissions 
traversing short terrestrial links (using HTTPS). Nevertheless, and depending 
on the TLS interception approach, the satellite link may have to be used during the
    TLS handshake. Figure~\ref{results2} shows the time required to load a simple
text-based Web page using HTTPS. In this figure, the bars corresponding to 
the case where Encrypted Content-Encoding (ECE) is used (i.e., the application
    layer encryption approach), the page is retrieved from a cache,
    otherwise it is retrieved directly from the server.

Additional advantages that can be gained, include: 
    (i) if terminals are connected using a terrestrial network, collaborative caching
solutions can be deployed, (ii) popular content (such as OS updates) can be pushed
to terminals, (iii) simultaneous content requests can be batched or even aggregated, 
(iv) solutions that achieve better performance by manipulating the transmitted content
(such as network coding) can be easily deployed. Quantifying 
these advantages has been left as future work.

\subsection{Security considerations}
The presented solutions are more secure compared to the certificate-based solutions
used by CDNs. With the solutions presented in this paper TLS connections are intercepted only for a specific session (with Keyless TLS), or
only for a specific period of time (with DANE with delegation semantics). Furthermore,
with the presented solutions it is easier for a content provider to revoke the
access rights of an intercepting device.
Nevertheless, TLS splitting and the use of application layer encryption security solutions
create some security and privacy concerns. 

Intercepting devices are authorized to 
access content items stored under a specific domain, hence they can easily
modify them. The integrity of the received items should be verified at the application
layer using tools such as the ``Subresource Integrity'' HTML tag. Furthermore,
application designers should make sure that sensitive
content, including user specific information, session identifiers, cookies,
    etc. cannot be accessed by third parties, even if the TLS session is intercepted.
    One should make sure that sensitive information and less sensitive content items are
stored under different domains.

When it comes to DANE with delegation semantics, there is a time frame during
which a ``de-authorized'' device can intercept a TLS session.
    This duration depends on 
the time-to-live (TTL) of the corresponding TLSA record. 
Application designers should consider this performance-security trade-off
and adapt TTL accordingly. 
     Note that there is also such a time frame with Keyless TLS,
    related to the TTL of a TLS session, but in this case
    ``de-authorized'' devices can only intercept already established sessions
and not new ones.

    Application layer encryption on the other hand creates privacy concerns.
Since two HTTP messages,
between two entities sharing the same key,
    are encrypting the same (secret) information,
it is possible for a malicious user that observes the network
    traffic to discern if the content of these messages is the same or not.
Furthermore, all network fields below the application layer are transmitted
in plaintext. Application designers should take special precautions in order
to properly anonymize transmitted messages.

\section{Conclusions}
Although end-to-end encryption enhances greatly
    end-user security and privacy, there are
cases where access to the plaintext of the transmitted content by in-network devices is beneficial. 
For this reason, various solutions for 
enabling encrypted connection interception have been proposed. In this paper, and
in the context of integrated satellite-terrestrial networks, we considered two of them, namely
Keyless TLS and DANE with delegation semantics. The former solution is ``pushed'' by big CDN
providers and requires no modification to TLS clients, whereas the latter solution is based
on a promising standard and it can achieve TLS session establishment without any 
communication with the origin server. Both solutions can be deployed using readily
available software and can co-exist with legacy TLS implementations. Furthermore,
both solutions exhibit better security properties compared to the certificate-based
solutions currently used by CDN providers.

The presented transport layer solutions can be combined with application layer encryption between
a satellite terminal and the origin TLS server (i.e., the network path that includes the satellite network).
The combination of the two approaches, creates interesting opportunities. In particular by sharing the 
application layer encryption key among the application server and the satellite terminals, it becomes possible
to broadcast content to multiple terminals, enabling this way existing satellite-based content
distribution solutions to operate with TLS 
    clients and facilitating the deployment of
novel optimization solutions, such as opportunistic caching.  

\section*{Acknowledgment}
Partial support for this work has been received by  
ESA
ARTES FP Project ``SATNEX-IV COO2.''
This paper does not necessarily reflect ESA views.

The work has been performed while Luca Boero was employed at the University of 
     Genoa.


\bibliographystyle{IEEEtran}
\bibliography{globecom}


\end{document}